# "FabSearch": A 3D CAD Model Based Search Engine for Sourcing Manufacturing Services

**Atin Angrish**
Edward P. Fitts Department of Industrial and Systems Engineering
North Carolina State University
111 Lampe Dr, Raleigh, NC 27695
aangris@ncsu.edu

**Benjamin Craver**
Edward P. Fitts Department of Industrial and Systems Engineering
North Carolina State University
111 Lampe Dr, Raleigh, NC 27695
bhcraver@ncsu.edu

**Binil Starly**
Edward P. Fitts Department of Industrial and Systems Engineering
North Carolina State University
111 Lampe Dr, Raleigh, NC 27695
bstarly@ncsu.edu

**ABSTRACT**

*In this paper, we present "FabSearch", a prototype search engine for sourcing manufacturer service providers, by making use of the product manufacturing information contained within a 3D digital file of a product. FabSearch is designed to take in a query 3D model, such as the .STEP file of a part model which then produces a ranked list of job shop service providers who are best suited to fabricate the part. Service providers may have potentially built hundreds to thousands of parts with associated part 3D models over time. FabSearch assumes that these service providers have shared shape signatures of the part models built previously to enable the algorithm to most effectively rank the service providers who have the most experience to build the query part model. FabSearch has two*

*important features that helps it produce relevant results. First, it makes use of the shape characteristics of the 3D part by calculating the Spherical Harmonics signature of the part to calculate the most similar shapes built previously be job shop service providers. Second, FabSearch utilizes meta-data about each part, such as material specification, tolerance requirements to help improve the search results based on the specific query model requirements. The algorithm is tested against a repository containing more than 2000 models distributed across various job shop service providers. For the first time, we show the potential for utilizing the rich information contained within a 3D part model to automate the sourcing and eventual selection of manufacturing service providers.*

1. INTRODUCTION

Advancements in product design software has expanded the user base of those involved in the design of new products. The easy availability of cloud-based design and manufacturing software and hardware has lowered the barriers of technical skill required to partake in the development process [1]. However, this ease of product design has not translated to providing tools for finding manufacturers that are best suited to fabricate prototypes or those that can conduct small production run quantities of new product systems. In large design enterprises, finding capable and reliable discrete part manufacturers is time consuming and is often tasked to sourcing agents who scour through existing supplier databases either within the enterprise or through known trust relationships established within the supply chain. The trust networks among the tiered supply chain for a product system are highly calcified and commercial ties remain in existence only because of historical reasons and risk aversion of decision makers and vendor selectors [2]. Sourcing, selection and compliance checks for suppliers are time consuming expensive activities but finding new sourcing partners are critical to maintaining the competitive edge of large and small enterprises. Much of the sourcing, finding, vetting and selection of new service suppliers is manual and hardly automatic. Manual processes and human trust based relationships prevent the easy formation and dissolution of supply chain service providers necessary for the production of one-off parts or personalized products since suppliers may always need to constantly change due to the need for reduced cost and faster delivery times.

There are more than 250,000 small manufacturer enterprises (SME) in the US with about 30,000 of them in machine shop job services [3]. From an SME service provider's point of view, finding and retaining new customers for the machine shop services is critical to business sustenance. Many small manufacturers spend a significant amount of their budget on marketing expenses to project their capabilities in front of existing and potentially new customers. This projection of capabilities is made possible by being constantly engaged with their clients through trade shows, establishing a sales network, digital advertising strategies and close personal contacts. These job shops are often small with less than a 100 people working on the shop-floor. Very few of them have business development teams that search and acquire new business contracts. Many of the websites utilized by these SME are rudimentary and do not possess the latest web technology to allow them to be competitive in connecting with web savvy young product designers. Once a product designer does find a manufacturer, it would take multiple communication rounds to determine if there is a good technical fit between designers' requirements and manufacturer capabilities. Currently, there is no automated tool to find the capabilities of manufacturers. Utilizing generic keyword based search engines, such as through Google, does not yield relevant results since indexing by search algorithms are based on text content on websites and how often they are cited. Listing services such as ThomasNet® provides semantic search capabilities, provided that a user knows the kind of manufacturing process that he/she is looking for [4]. Often the rankings of the manufacturers from keyword lookups are not transparent and often proportional to the advertising expenses spent by the SME in order to rise up in search results in these web based platforms.

In this paper, we propose "FabSearch", a prototype to connect manufacturing service providers with users who are searching for their services by having them conduct a 3D model based search query to find manufacturers. The algorithm would generate a ranked list of service providers who are best suited to fabricate the part. The underlying assumption is that finding the best manufacturing process service providers for a particular query part design model are the ones who have made similar parts to the query part with comparable material specifications and tolerance requirements in the past. If we assume that service providers are willing to share specific shape signatures of part models they have built in the past, our algorithm captures the shape signature of the 3D model query based on information contained with its definition, and then compares it against a repository of part models built previously by the manufacturing service provider. This concept would be a shift away from how new manufacturing service suppliers are found and sourced. The design of the part and information contained within it drives the supplier search process as opposed to selecting service providers through human trust based relationships. The remainder of this paper describes the process through which a ranked list of supplier services best suited to make a particular query design part is generated. We test our algorithm against more than 2000 product models assumed to be shared by 13 service providers with various process capabilities.

## 2. RELATED WORK

One of the most important operational tasks in order to ensure smooth and reliable production of any product assembly is the optimal identification of manufacturers given numerous constraints ranging from historic performance, quality rating, certification, geographic location, experience, cost etc. As a result, a large body of research has been performed with regards to optimal manufacturer selection across various product category domains of aerospace, automobile, textiles, semiconductor industry. The optimality of choice often depends on several constraints such as manufacturer competence, experience, trust level and the management structure with the values of these metrics often subjectively decided by the team conducting the pruning and selection process. Since the problem of supplier selection is often a multidimensional one involving several stakeholders, the most popular methodologies for solving them have been the techniques used in multi criteria decision making (MCDM) tools such as Analytical Hierarchy Process (AHP), Analytical Network Process (ANP), Fuzzy AHP (F-AHP), TOPSIS etc. Relevant work by Noci et al [5] and Lee et al [6] develop a green supplier selection in different industry types by the use of AHP and ANP methods. Similarly, there have been several works extending the AHP/ANP methodologies using fuzzy techniques by Kahraman et al [7], Kilincci and Onal [8] and Chamodrakas et al [9] for identification of suppliers for the manufacturing of consumer products in electronic marketplaces. Similarly, extensive work has been done using methodologies such as TOPSIS, as explained in the works by Boran et al [10], fuzzy hierarchical TOPSIS for generating supplier rankings by Wang, Cheng and Huang [11] and similar work by Liao and Kao [12]. A major limitation of these approaches is that these techniques often rely on subjective assessment of various characteristics of the vendors based on "expert" opinion, which can often be misleading and suffers from the lack of solution generalizability.

Another popular approach towards solving supplier selection problems under constraints has been the use of linear programming/optimization based methodologies. The main benefit of these techniques is that they allow a robust approach towards solving multi-objective problems with a given set of limitations. Often these are complimented by the use of genetic algorithms. Some of the more popular works in this area are by Xu and Ding [13] and Sadeghieh et al [14]. Liao and Rittscher also demonstrated the use of multi-objective supplier selection under stochastic demands [15]. Similar works extend the idea for solving the same problem under multi-period assembly constraints [16]. Approaches based on linear multi-objective optimization are more robust mathematically compared to MCDM methods. However, they also suffer from the problem of subjectivity in their initial assumptions used to drive the selection process.

A number of academic papers have made use of a multitude of parameters pertinent to manufacturing such as energy consumption, transportation costs, inventory management etc. to drive the supplier service selection process in discrete manufacturing. Such metrics are relatively easily obtained or may be simulated with appropriate distributions based on historical data. At the same time, data based on subjective criteria such as quality of past work, trust levels and product design requirements and historical manufacturer performance are very hard to obtain and can often vary based on the team conducting the selection. As a result, to the best of the authors' knowledge, no papers have been published which utilize the product manufacturing information (PMI) such as the computer aided design models (CAD models), tolerance requirements and the material requirements for a given design to search suitable manufacturers for the part design who have the best expertise and experience needed to fabricate the part.

Search algorithms involving the use of 3D model surface mesh have been extensively researched, primarily from the perspective of retrieving similarly shaped parts from a database. Iyer et. al. [17] divides major methods of shape search into major categories: Manufacturing feature recognition (feature relationship graphs [18]), global feature descriptors [19][20], local feature descriptors [21], graph based methodologies [22] and Histogram based [23][24][25] techniques. As discussed in the paper, shape search algorithms can have a variety of applications beyond the areas of computer graphics when combined with appropriate databases and tacit domain knowledge, these applications can find numerous applications in manufacturing. Recent advances in computer vision techniques to search for complex shapes are enabled by the use of Extended Gaussian Images [26] and Convolutional Neural Networks [27][28] which often rely on supervised learning methods for classification and retrieval tasks. However this is often achieved at high computational costs and with large amounts of training data involved to train these networks. These methods are not directly applicable to solving the problem of manufacturer search which relies not just on the shape of the 3D part, but also on related process specifications such as assigned material, geometric and dimensional tolerance requirements on part features that plays a direct role on the process planning and appropriate suppliers who can adhere to such specifications. There has been no reported work on the use of shape search methodologies for supplier selection in manufacturing.

## 3. "FABSEARCH" SYSTEM OVERVIEW AND METHODOLOGY

It is fairly common knowledge in manufacturing that a completely new project for a service provider based on new client requests requires significant planning and investment of time and effort by the manufacturer in order to procure appropriate material, build necessary competency and tooling for the successful execution of various job orders. Hence, service providers tend to mostly accept job orders with which they have most experience in, even if they are fully capable of producing parts outside of work they have done in the past. However, users will not be able to automatically seek and find such experienced manufacturers. If such users had the ability to search for service providers who have done similar jobs, product development times can be shortened yielding better quality results and faster product lifecycle times. Hence, the central paradigm for the development of a manufacturing search engine is the following: Manufacturers who have worked with similar parts in the past are more "suitable" for making a new part of the product assembly system, since they have the tacit knowledge and the tools available for the new projects based on experience and data available from old projects.

A corollary to this premise is that parts with similar features and process requirements such as tolerance, material may be made by the same manufacturers. Using the CAD data of the part provided by the search engine user and searched against data provided by the manufacturing service provider's past work, FabSearch allows us to generate a ranking for the suitable manufacturers. For example, a service provider may already have all of the process planning steps required to machine a part on a CNC machine. If a user is able to find such service providers based on similarity of shapes and GD&T requirements, the service provider need only modify older designs to recreate the process toolpath needed for the new part, thus saving significant time and costs involved in the new project. An overview of the search engine process is presented in Fig 1.

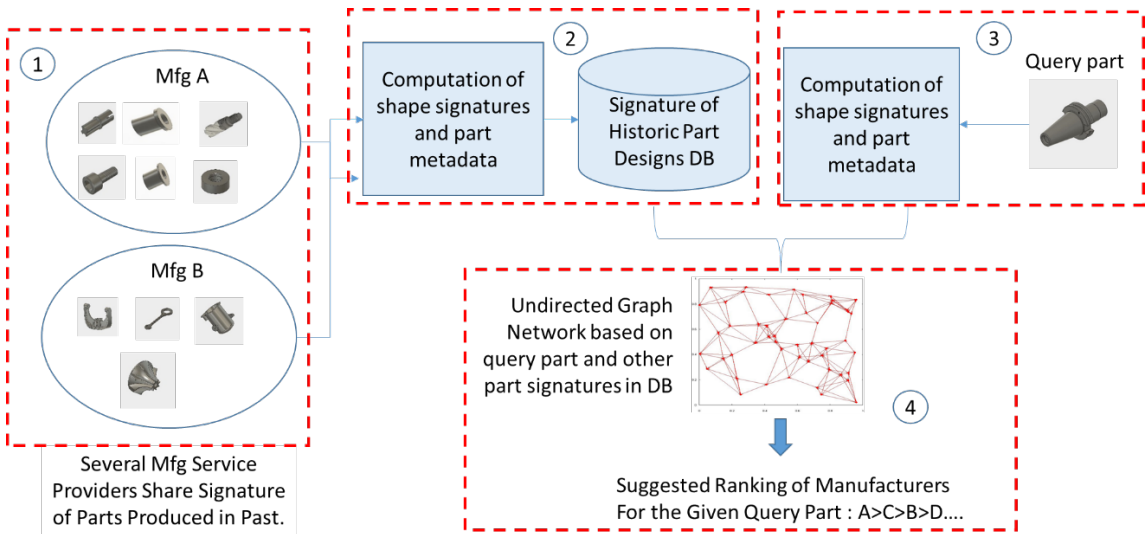

*Figure 1: Manufacturing search engine matching query part to historic parts produced by manufacturers*

Assuming that if the manufacturers share signatures of CAD data in the form of .STEP files of the parts from their previous projects in Fig 1(subset1), the search engine can compute a shape signature for the CAD model with the part specific metadata and stores the information in a database (subset 2). When a designer needs to make a part, they upload their own CAD model to the engine, which calculates the signatures for the query model and then compares the signatures against the database which then results in the ranking of manufacturers based on their past experience of making similarly shaped parts with similar material assignments and GD&T specifications (sub-set Region 3 and 4 of Figure 1).

In the following sections, the methodology used for generation of a suitable dataset for simulation of the scenario, the explanation for the algorithms used have been described. A case study on how the algorithm generates a ranking for suppliers for our proposed problem statement has also been presented.  The use of a shape signature has also been proposed to be used by a search engine which would query a database compiled from the CAD model shared voluntarily by the manufacturers for the parts that have been made by them in the past. The shape signature shall allow us to build an undirected K nearest neighbor graph search space followed by a Bayesian ranking algorithm.

### 3.1. Development of 3D Part Model Dataset – "FabWave Repository"

To adequately test the search engine algorithm, a large number of 3D engineering definition of 3D part models was required. There are a number of shape based repositories available online [29][30][31][32] which contain different kinds of shapes used by the computer graphics community for the development, validation and testing of shape classification and retrieval algorithms. However, data contained in these repositories contain 3D models of assemblies with incomplete information necessary to fabricate the part. For example, the models in the repository contain surface mesh data without the necessary boundary representation (B-REP) information required to fully define part models. The most popular repository for engineering parts - Engineering Shape Benchmark(ESB) [30] was developed by Jayanti and Ramani et. al with about 867 engineering oriented models. The repository contained models belonging to 3 main categories: Flat-thin wall components, Rectangular-cubic prism and solids of revolution, with several varying components in each of the three categories. The repository still lacked sufficient number of models with data only stored in the .STL and .OBJ formats therefore losing all feature level and associated meta-data (units, material, geometric tolerance) necessary for fabrication.

To test our algorithm, we built up a repository of 3D part models, called "FabWave" through several methods. In the first approach, we crawled the web and downloaded part assemblies from community generated data such as those available through the Autodesk Fusion Gallery and GrabCAD. Only those models that have been explicitly shared by the original author was included in the automated crawl and retrieval. In the second approach, we built a template model for standard parts (ex. washers, rings, bolts, seals etc.) of a part and iterated through parametric variations on the features based on standard dimensions obtained from websites of part supply companies such as McMaster-Carr and Global Spec. As of Sept 2018, we have collected more than 100,000 engineering part models in either their native CAD format (such as Fusion .f3D) or in the cross platform .STEP format. The full repository is available for access at http://www.dimelab.org/fabwave [33].

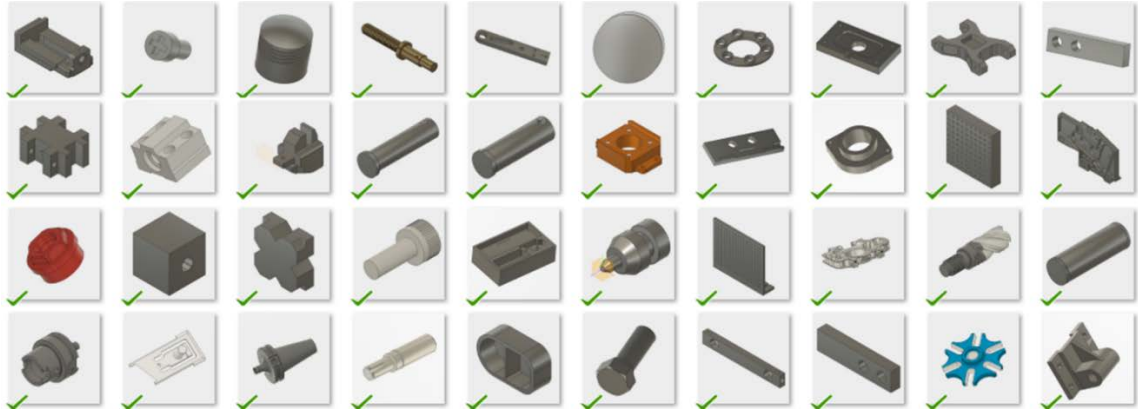
*Figure 2: Sample of non-standard components*

For the purposes of testing our approach, a subset of 2133 models from the entire dataset was divided into two major classes: standard parts and non-standard parts. The standard parts in the repository contain standardized models such as washers, nuts, bolts, springs etc. The non-standard parts contain freeform and prismatic 3D models with sufficient nominal dimension specified that enable the part to be fabricated through traditional manufacturing processes. All models have their associated STEP representations containing part metadata and designer intent such as choice of material, dimensional units and tolerances associated with the parts. Every part generated within the repository has a 32 bit unique id assigned to ensure similar names were not generated. A related meta-data document was also generated for each product model such as the volume, surface area, volume/surface ratio, number of features. A sample set of the parts used in this study are as shown in Fig 2-3:

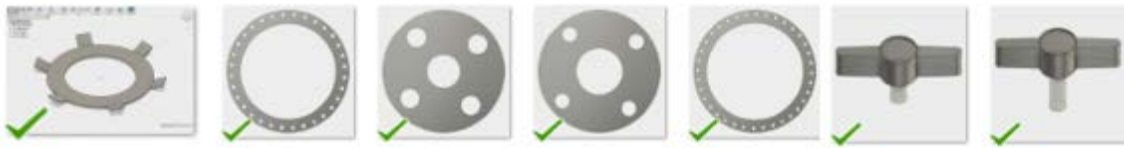
*Figure 3: Examples of standard parts (Screws, Washers, Sprockets, Brackets, and Retaining Rings)*

The models were segmented further with the following criteria:
1. Standard parts were broken down into subcategories as identified by conventional naming classification of these parts. Such sub-categories included brackets, washers, pipe fittings, o-rings, seals, springs, screws etc.
2. The Non-standard parts were assigned to a large unassigned group and were classified under manufacturing process categories as opposed to part name categories.

Since the original data for the non-standard 3D models was provided by the community, scripts were written to ensure critical missing information was added to all of the CAD models. This included adding in material assignment and including tolerance range assignments based on the process category a particular part was assigned to. For example, parts that were assigned to the machining category had assignments of tolerance callouts from high to the standard tolerance callouts. The following parameters were simulated for the models that did not have complete information:
- Material: Metals (Stainless Steel, Titanium, Fe, Cu) and Nonmetals (Plastics such as PP, Urethane etc.)

Insert ASME Journal Title in the Header Here- Tolerances: The simulation code was written to generate only the tightest dimensional tolerance for each part which was assigned using a sampling from an appropriate uniform distribution. The reasoning behind using only the tightest dimensional tolerance is the assumption that if the manufacturer can achieve the tightest tolerances, they are capable of producing parts with lower tolerances specifications as well. After the tolerances were assigned, K-Means clustering was used to segment them into 3 clusters: standard, medium and high tolerances.
- Manufacturing processes: Each part was assigned the manufacturing processes that would be required for making the parts with the required tolerances. While some parts may need multiple manufacturing processes, for the purposes of this study, only the "dominating" manufacturing process which will meet the majority of manufacturing tolerances required was considered. This was also used for further subdividing the non-standard parts into 4 process categories: Casting, Machining, Forming and Molding.
- Manufacturers for the part: To simulate the assumption that the manufacturer service provider has shared previous work, we randomly assigned parts belonging to a specific process category to unique manufacturers. The assignment was made in a manner such that the manufacturers have done almost equal number of parts, each in accordance to their own expertise. For example, a firm which primarily does casting was not assigned a part that requires the forming process. Each manufacturer was also assumed to do a single process category to test uniqueness of the solution.

The different breakups for the assignments and the repository are as shown in Table 1. Overall, there are 13 manufacturers with 2133 parts in the repository across 3 tolerance ranges and 16 categories (12 standard name categories and 4 non-standard process categories).

*Table 1: Manufacturer Service Provider Assignment by Process Category*

| Manufacturer | Casting | Forming | Machining | Molding | Grand Total |
|---|---|---|---|---|---|
| A | | 329 | | | 329 |
| B | | 334 | | | 334 |
| C | | 339 | | | 339 |
| D | | 346 | | 111 | 457 |
| E | | | | 96 | 96 |
| F | | | | 105 | 105 |
| G | | | 83 | 92 | 175 |
| H | | | 84 | | 84 |
| I | | | 76 | | 76 |
| J | 15 | | 81 | | 96 |
| K | 12 | | | | 12 |
| L | 15 | | | | 15 |
| M | 15 | | | | 15 |
| **Grand Total** | **57** | **1348** | **324** | **404** | **2133** |

Insert ASME Journal Title in the Header Here*Table 2: Part assignments by process*

| Parts | Casting | Forming | Machining | Molding | Grand Total |
|---|---|---|---|---|---|
| Bearings | | | 58 | | 58 |
| Bolts | | 10 | | | 10 |
| Brackets | | 54 | | | 54 |
| Casting | 57 | | | | 57 |
| Forming | | | | 46 | 46 |
| Machining | | | 255 | | 255 |
| Molding | | | | 67 | 67 |
| Other | | 78 | | | 78 |
| Pipe Fittings | | 33 | | | 33 |
| Retaining Rings | | 207 | | | 207 |
| Rollers | | | | 14 | 14 |
| Slotted Oval Head Screws | | 50 | | | 50 |
| Socket Head Screws | | 202 | | | 202 |
| Sprockets | | | | 277 | 277 |
| Unthreaded Flanges | | | 11 | | 11 |
| Washers | | 714 | | | 714 |
| **Grand Total** | **57** | **1348** | **324** | **404** | **2133** |

*Table 3: Part assignment by tolerance*

| | Tolerance Ranges | | | |
|---|---|---|---|---|
| Parts | High | Medium | Standard | Grand Total |
| Bearings | 58 | | | 58 |
| Bolts | | | 10 | 10 |
| Brackets | | | 54 | 54 |
| Casting | | 57 | | 57 |
| Forming | | 46 | | 46 |
| Machining | 255 | | | 255 |
| Molding | | 67 | | 67 |
| Other | | | 78 | 78 |
| Pipe Fittings | | | 33 | 33 |
| Retaining Rings | | | 207 | 207 |
| Rollers | | 14 | | 14 |
| Slotted Oval Head Screws | | | 50 | 50 |
| Socket Head Screws | | | 202 | 202 |
| Sprockets | | 277 | | 277 |
| Unthreaded Flanges | 11 | | | 11 |
| Washers | | | 714 | 714 |
| **Grand Total** | **324** | **461** | **1348** | **2133** |

*Table 4: Manufacturer assignment by part*

| Mfg Supplier | Bearings | Bolts | Brackets | Casting | Forming | Machining | Molding | Other | Pipe_Fittings | Retaining_Rings | Rollers | Slotted_screws | Socket_screws | Sprodkets | Flanges | Washers | Total |
|---|---|---|---|---|---|---|---|---|---|---|---|---|---|---|---|---|---|
| A |  | 3 | 19 |  |  |  | 17 | 10 | 49 |  | 11 | 39 |  |  | 181 |  | 329 |
| B |  | 1 | 16 |  |  |  | 22 | 8 | 51 |  | 16 | 48 |  |  | 172 |  | 334 |
| C |  | 1 | 10 |  |  |  | 24 | 9 | 50 |  | 9 | 69 |  |  | 167 |  | 339 |
| D |  | 5 | 9 | 16 |  | 19 | 15 | 6 | 57 | 5 | 14 | 46 | 71 |  | 194 |  | 457 |
| E |  |  |  | 10 |  | 16 |  |  |  | 1 |  |  | 69 |  |  |  | 96 |
| F |  |  |  | 12 |  | 20 |  |  |  | 6 |  |  | 67 |  |  |  | 105 |
| G | 16 |  |  | 8 | 66 | 12 |  |  |  | 2 |  |  | 70 | 1 |  |  | 175 |
| H | 14 |  |  |  | 65 |  |  |  |  |  |  |  |  | 5 |  |  | 84 |
| I | 9 |  |  |  | 65 |  |  |  |  |  |  |  |  | 2 |  |  | 76 |
| J | 19 |  |  | 15 | 59 |  |  |  |  |  |  |  |  | 3 |  |  | 96 |
| K |  |  |  | 12 |  |  |  |  |  |  |  |  |  |  |  |  | 12 |
| L |  |  |  | 15 |  |  |  |  |  |  |  |  |  |  |  |  | 15 |
| M |  |  |  | 15 |  |  |  |  |  |  |  |  |  |  |  |  | 15 |
| Total | 58 | 10 | 54 | 57 | 46 | 255 | 67 | 78 | 33 | 207 | 14 | 50 | 202 | 277 | 11 | 714 | 2133 |

## 3.2. Shape Analysis methodology:

Spherical Harmonics [25] was used for our implementation to capture the shape signature of the part. This descriptor was designed for rigid body search which is appropriate for the rigid parts contained within the dataset. It is an established and a well-known global shape descriptor, inexpensive to compute (~0.28s on average on the FabWave dataset), has good discriminative ability (~65% on the Princeton Shape Benchmark) and is deterministic in nature. The SPH also allows us to compare shapes using simple Euclidean distance between the query signature and signatures of shapes within the search space. A quick overview of the SPH algorithm is as follows:

1. Voxelize the model to R x R x R grid, where R is a whole positive number that allows the model to be sufficiently voxelized to gather enough detail necessary. The value of R, typically ranges from ($2^5 - 2^8$). Assigned value of R=32.
2. Translate voxelized model to (R/2,R/2,R/2) and scale to size R
3. Convert voxel center coordinates to spherical coordinates with radius R and angles $\theta$ and $\phi$
4. Discretize the radius to n concentric spheres and compute the following:

$$f_r(\theta, \phi) = \sum_m f_r^m(\theta, \phi) \text{ where}$$

$$f_r^m(\theta, \phi) = \sum_{n=-m}^{m} a_{mn} P_{mn}(\cos\theta) e^{in\phi} \sqrt{\frac{(2m+1)(m-|n|)!}{4\pi (m+|n|)!}} \dots\dots\dots(2)$$

, where $P_{mn}$ is Legendre polynomial. This generates a n x m signature unique for each part.

The n x m signatures provides an easy way to compare the shape signatures of different parts. The models in the repository were processed through the SPH algorithm to find their respective signatures for each part. The difference between a query model (denoted Q) and the models in the repository ( $r_i \in R$ ) , where R denotes the repository set and $r_i$ denotes the i-th model in the repository) by comparing the $L_2$ norm difference of the signatures. The difference may be treated as a distance between the models in multi-dimensional space and therefore maybe used to build a K-nearest neighbor graph [34].

### 3.3. Backlink Generation

While it is possible for us to generate a robust ranking based on a simple K-nearest neighbor algorithm using KD-trees, the decision on the appropriate value of K would yield subjective results. To circumvent the issue of subjectivity and to generate a comprehensive ranking system, we use a technique inspired by the PageRank algorithm [35]. A graph is built with each node containing the mapping $\rightarrow \{S, M, T\}$ : where P contains the part ID and the triplet $\{S, M, T\}$ contains the ordered set of the shape signature computed from the SPH method and M denotes the manufacturer made with tolerance T. The $S$ component for $Q$ is used for generation of a KNN graph. At the same time, all the K-nearest neighbors for all the existing parts in the repository can be found. This technique is similar to finding backlinks from all webpages to an existing page in the PageRank algorithm. This allows us to find not only the parts $Q \sim r_{i,s}$ but also $r_{i,s} \sim Q$. This may be thought of as finding KNN graph for all models including $r_{i,s}$ and $Q$ and converting the directed graph thus generated into an undirected graph. As a result, a more relevant search space for a query part can be found. This process is illustrated in the figure below:

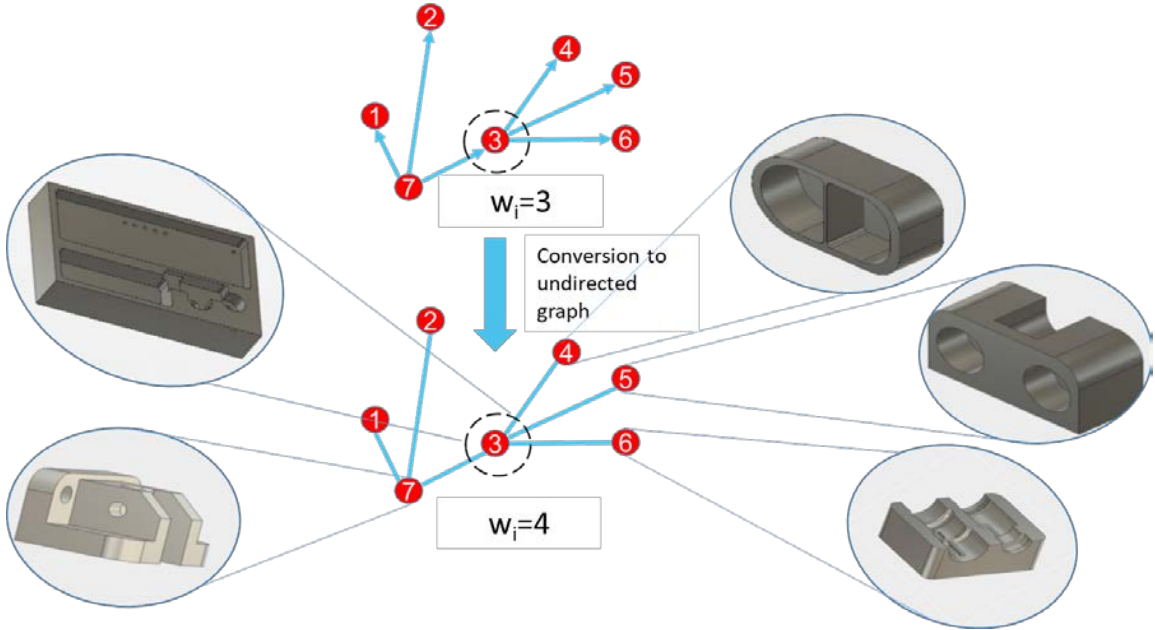

*Figure 4: Conversion of directed kNN graphs to undirected graph (Part ID 3 is the Query part)*

The above methodology also enables parts which are mutually similar to each other to be clustered together. Since the nodes of the graph contain the manufacturer information as well, consequently having the "nearest" manufacturers for part Q may be found as well. The nodes contain information about the manufacturer M and the tolerance T, which may then be used to rank the manufacturers using the Bayes Theorem:

$$P(M = m \mid (S \sim S_i \in \{K_{i \to j} \cup K_{j \to i}\} \cap T = x \in \{L, M, H\})) \quad \ldots\ldots. (3)$$

Equation (3) calculates the probability of a given manufacturer 'm' belonging to the nodes connected to the query part Q (having shape signature S). This is matched from the bidirectional KNN graph (denoted by $K_{a \to b}$, where $a \to b$ signifying $b$ is in the neighborhood of $a$) which find similar shapes with tolerances which match the requirement of the search part. The ranked solution would then approximate how close any manufacturer is suited to making the query part, based on historical parts built by each manufacturer service company in the past. The final ranking is sorted in a descending order to generate a list of manufacturers based on a descending order of their capability of finishing the query job fulfilling all of the PMI requirements.

## 4. RESULTS

The search engine was tested against the following query parts (Fig 5-8). Each of these parts contain holes, freeform shaped pockets and multiple intersecting cut features that can be a challenge for any shape based engine search. For the first query part, the original 3D part model information along with associated material search requirement equal to metal alloy and a 'HIGH' tolerance requirement was set as the search criteria. When the part is processed through the proposed engine, the part is voxelized and SPH signatures are generated for the query model. These signatures are used to find the $L_2$ distance from all the parts in the library and vice versa, using the modification of the KNN graphs. As a result, the following results are observed for query part 1:

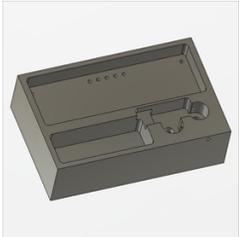

*Figure 5: Sample Query Part and the results (circled shapes indicate the closest shape match)*

In the Fig 5 query search, it is noted that the parts circled in red are quite similar to the query part in shape, such as presence of two pockets, being "blocky" in shape and with holes. The results are also tagged with the tolerance assignment of the parts and the associated manufacturer who was assigned to making those parts. From Eqn (3), the ranking for the suggested manufacturers would then be H > A > J = M = L = I > C. False positives were also reported by the algorithm. However, it is observed that for the 10 nearest neighbor graphs, the outcomes of the undirected graphs with backlinks lie between 12 – 100 results for the dataset that we tested the query against. Other sample parts with the associated results and the suggested manufacturer rankings are as shown in the Figures (Fig 6-8):

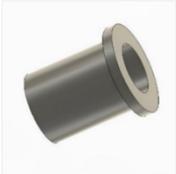

*Figure 6: Sample Query 2*

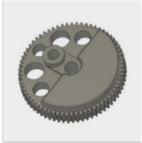

Figure 7: Sample Query 3

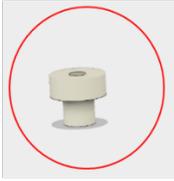

Figure 8: Sample Query 4

Based on the fact that the engine assigned plastic parts to injection molding, 2033 metal parts were left from the repository. The metal parts are categorized into their dominant manufacturing processes and the manufacturers that were assigned to them in the simulation. Each part in the metals category is then run through the search engine and the following two metrics were used to evaluate the search results:

- Metric 1 – Correct Manufacturer Type Assignment: Within each category of manufacturing processes, in how many cases was a part assigned to a manufacturer who specializes in the correct manufacturing process as required for the part. For example, if the dominant manufacturing process for the query part is machining, the part should be assigned to a service provider who specializes in machining (who may or may not be the original vendor assigned at the beginning of the simulation)

- Metric 2 – Original vs New Manufacturer Assignment: If Metric 1 is satisfied (i.e. For the cases where the correct assignments were made), in how many cases was an improved assignment made compared to the one originally given manufacturer for the query part. That means if a part was made by a manufacturer X but the search engine finds a manufacturer Y who has done more similar parts with similar materials and tolerance requirement, the recommended vendor from the engine is vendor Y (which would have been a better assignment compared to X, which was made without any priori information about the vendor).

The search engine was evaluated based on these metrics and the results are as follows:

*Table 5: Search Engine performance on Metric 1*

| Category | Number of parts | Number of correct assignments | Percentage correct assignments |
|---|---|---|---|
| Casting | 57 | 14 | 25% |
| Forming | 1348 | 1320 | 98% |
| Machining | 324 | 243 | 75% |
| Molding(Powder Metallurgy) | 2066 | 1870 | 87% |
| Total | 2066 | 1870 | 87% |

On metric 1, the search engine shows that it is able to identify the correct manufacturers for a given part almost 87% of the time. The worst performance is with the casting part manufacturers who were often misidentified for machining. Given the complexity and the similarity of the parts in the casting and the machining dataset, this problem was expected. Also, our dataset was very small for casting parts (n = 57). The engine performed fairly well on the remaining 3 categories of manufacturing processes. Molding with metals implies powder metallurgy parts. The engine performs fairly well with forming followed by molding and machining parts respectively. This could be due the variability of the parts in the datasets. Often times, formed parts are fairly simple designs with an approximately uniform distribution across sizes and shapes. Same goes with parts such as gears and handles in the powder metallurgy / molding category. The machining dataset has the most variability in designs compared to any other dataset in the repository containing parts with low to high complexity of features within the part. In 25% cases, the parts have been matched to forming and casting manufacturers due to the similarity of designs.

Insert ASME Journal Title in the Header Here*Table 6: Search Engine Performance on Metric 2*

| Category | Improved assignments in correctly identified parts | Improved assignments as a percent of all parts |
|---|---|---|
| **Casting** | 50% | 12.5% |
| **Forming** | 23.4% | 22.9% |
| **Machining** | 51.4% | 38.4% |
| **Molding / Powder Metallurgy** | 35.1% | 30.8% |
| **Total** | 40.0% | 34.8% |

As seen in Table 6, the search engine allows us to identify better manufacturers in all categories of parts present in our repository. This is ensured due to the use of Bayesian ranking and the large search space created by the back-linking process in our dataset. It can be seen that there is a significant improvement in the number of improved assignments in all categories except for the casting dataset, which as noted earlier suffers from the problem of a small dataset within the category. It is noted that significant benefits have been observed in the datasets with large number of models. This implies that it is possible to have better results with a repository that contains a larger number of diverse models.

## 5. DISCUSSION

The proposed methodology gives a suitable way for search and evaluation of manufacturers based on their past experience with different CAD models based on designer part requirements. The primary contributions of this paper is the development of an appropriate dataset (available for use on request), use of back linked nearest neighbor graphs and the Bayesian ranking for manufacturers based on CAD and metadata similarity, allowing us to find potentially better manufacturers experienced with projects similar to the query parts. As we can see, the approach benefits from large numbers of models shared by manufacturers with as many different diverse models. The larger the variety of shapes and associated metadata shared by manufacturers, the better the search engine would perform. It is crucial to note that with continuous updating and sharing of the models by the manufacturers, the manufacturer rankings may also change with time. The large amounts of specialized knowledge in the form of experiential learning by the manufacturers might be difficult to harness by machine learning methodology. However, we can rely on their past jobs to project the potential success of working with a similar but new project. Empirically, it can be expected that the cost for the new project execution will be lower because the manufacturer already has the knowledge and possibly even the tooling, material and the experience required for making the part. This can lower the barriers to product development lifecycles by improving the efficiency through which sourcing is conducted.

The authors note that there are several limitations to the approach taken in the paper. The first limitation of the engine is the shape search algorithm used. While SPH is a powerful shape search methodology, the performance of the algorithm has been surpassed by newer methodologies such as multi-view convolutional neural networks [28], OctNets [36],

VoxNets [27] etc. However, these newer shape search algorithms requires significant computational power and tremendous amount of training on various classified models to allow deep learning algorithms to perform equally well. Another point of note is that the KNN generation uses a brute force approach to distance calculation which might not scale well as the size of the shared repository increases. Newer approaches towards KNN graph generation using approaches in [37][38][39][40] may be used to speed up the computation as the number of models in the repository increases.

A potential upgrade to the search engine could be the use of additional metadata such as the size of the part, use of the rich annotated text data available with associated PMI of the CAD models. The search engine results are also dependent on the amount of data shared by the individual manufacturing service companies. The search engine should also be able to allow multi-modal query data to be entered, i.e. 3D models with associated text and possibly even images to improve the relevancy of the search results. The search engine should be able to incorporate information from the content available on the manufacturing service providers' websites to help reach an optimal selection for a designer searching for appropriate manufacturing services for their designs. A possible future version of the search engine could rely on joint embedding of multimodal data such as shape, tolerance and text information into a single higher dimensional manifold by use of neural networks with the objective function to reduce the distance between similar models [41] and [42].

Sharing of part model information by manufacturers themselves could be of concern. Even if service providers may be willing to share the model themselves, legal contracts bind them from not directly sharing the raw 3D CAD model with an external third party without the explicit permission of the original client. Our current implementation does not require the service provider to directly share the original 3D model. Rather, only a signature of the part model need be shared with a third-party to increase the chances of the service provider being found. This aspect can perhaps lower the barrier of sharing past work done by a particular service provider, given that the original 3D model cannot be reverse engineered from the obtained signature. If the manufacturers did share the signature of CAD models of previous activities, it is observed that our algorithm is inherently biased towards manufacturers who share more data than those who do not which means deserving manufacturers might not be able to get equal opportunities. However, this is no different than a typical business which lags behind in Google's search results due to an ineffective Search Engine Optimization (SEO) approach for a particular vendor's website. To tackle the issue of computing on encrypted data without directly sharing the original files with an external third party is through the use of homomorphic encryption schemes [43] which allow rudimentary computations on encrypted data. Appropriate incentivization mechanisms can also be devised to encourage job shop service providers to share part model data.

## 6. CONCLUSIONS
This paper has discussed the paradigm of a manufacturing search engine which utilizes the CAD PMI for recommendations of manufacturers based on a query 3D model and associated manufacturing requirements. The development of such an engine required the development of a custom repository of manufacturing components, the methods of collection and simulation for the same were described in this paper. Also, this paper discussed the use of shape harmonics for evaluation of shape similarity followed by development of kNN graphs with backlinks for generation of a Bayesian ranking methodology for manufacturer selection based on similarity of past work to the query models. Further development in this field is possible by borrowing ideas from computer vision, data encryption and manufacturing fields for making a more expansive search engine solution that yield quality results.

## 7. ACKNOWLEDGMENT
The authors would like to thank the anonymous reviewers for their valuable comments.

## 8. FUNDING
This study was supported by a grant from the US National Science Foundation (NSF-CMMI 1547105 and OAC-